\documentclass[final,5p,times,twocolumn,sort&compress]{elsarticle}

\usepackage{amsmath}
\usepackage{amssymb}
\usepackage{natbib}
\usepackage{graphicx}
\usepackage[version=3]{mhchem}
\usepackage{physics}
\usepackage{url}
\usepackage{color}
\usepackage{appendix}

\newcommand*\dif{\mathop{}\!\mathrm{d}}
\newcommand*{\onlinecite}[1]{\hspace{-1 ex} \nocite{#1}\citenum{#1}}
\DeclareMathOperator{\sgn}{sgn}

\newcommand{\myfig}[4][ht]{
\begin{figure}[#1]
\centering
\includegraphics[#2]{#3}
\caption{#4\label{#3}}
\end{figure}
}

\newcommand{\myfigwide}[4][ht]{
\begin{figure*}[#1]
\centering
\includegraphics[#2]{#3}
\caption{#4\label{#3}}
\end{figure*}
}

\graphicspath{{images/}}

\newcounter{bla}

\journal{Computer Physics Communications}
\begin{document}
\begin{frontmatter}

\title{\texttt{almaBTE}: a solver of the space-time dependent Boltzmann\\transport equation for phonons in structured materials}
\author[cea,tuw]{Jes\'us Carrete\corref{cor1}}
\author[cea]{Bjorn Vermeersch\corref{cor2}}
\author[cea]{Ankita Katre}
\author[cea]{Ambroise van Roekeghem}
\author[rub]{Tao Wang}
\author[tuw]{Georg K. H. Madsen}
\author[cea]{Natalio Mingo\corref{cor3}}

\cortext[cor1]{Corresponding author.\\\textit{E-mail address:} jesus.carrete.montana@tuwien.ac.at}
\cortext[cor2]{Corresponding author.\\\textit{E-mail address:} bjorn.vermeersch@cea.fr}
\cortext[cor3]{Corresponding author.\\\textit{E-mail address:} natalio.mingo@cea.fr}
\address[cea]{Universit\'e Grenoble Alpes, F-38000 Grenoble, France\\
  CEA, LITEN, 17 rue des Martyrs, F-38054 Grenoble, France}
\address[tuw]{Institute of Materials Chemistry, TU Wien, A-1060 Vienna, Austria}
\address[rub]{CMAT, ICAMS, Ruhr-Universität Bochum, 44780 Bochum, Germany}

\begin{abstract}
\texttt{almaBTE} is a software package that solves the space- and time-dependent Boltzmann transport equation for phonons, using only \textit{ab-initio} calculated quantities as inputs. The program can predictively tackle phonon transport in bulk crystals and alloys, thin films, superlattices, and multiscale structures with size features in the $\mathrm{nm}$-$\mathrm{\mu m}$ range. Among many other quantities, the program can output thermal conductances and effective thermal conductivities, space-resolved average temperature profiles, and heat-current distributions resolved in frequency and space. Its first-principles character makes \texttt{almaBTE} especially well suited to investigate novel materials and structures. This article gives an overview of the program structure and presents illustrative examples for some of its uses.

\end{abstract}

\end{frontmatter}

{\bf PROGRAM SUMMARY}

\begin{small}
  \noindent%\hspace{0.5em}
      {\em Program Title:} \texttt{almaBTE}\\
      {\em Journal Reference:}\\
      {\em Catalogue identifier:}\\
      {\em Licensing provisions:} Apache License, version 2.0\\
      {\em Programming language:} C++\\
      {\em Computer:} non-specific\\
      {\em Operating system:} Linux, macOS\\
      {\em RAM:} up to tens of GB\\
      {\em Number of processors used:} variable
      {\em Classification:} 7.9 Transport Properties\\
      {\em External routines/libraries:} BOOST, MPI, Eigen, HDF5, spglib\\
      {\em Nature of problem:} Calculation of temperature profiles, thermal flux distributions and effective thermal conductivities in structured systems where heat is carried by phonons\\
      {\em Solution method:} Iterative solution of linearized phonon Boltzmann transport equation, Variance-reduced Monte Carlo\\
      {\em Running time:} Up to several hours on several tens of processor cores\\
\end{small}

\section{Introduction}

Thermal transport in semiconductor devices has become a growing priority in recent years. In many technologies, the problem of heat dissipation stands in the  way to further progress. Phonons being the main heat carriers in semiconductors and insulators, clearing the path requires tackling inherently multiscale problems from the point of view of the fundamental physics of vibrations in solids.

By way of example, in power electronic devices such as LEDs or High Electron Mobility Transistors, the choice of substrate is crucial for the lifetime of the device. Lowering the temperature of the active region by $50\,\mathrm{K}$ can increase a device's time to failure by one order of magnitude \cite{jimenez_x-band_2008}. Thus, various generations of multilayer structured substrates \cite{cahill_nanoscale_2014} have been evolving in recent times in order to increase heat flow, while also satisfying other constraints such as structural integrity and cost. The difficulty in this design process is that modeling thermal transport in these structures is much more complex than for macroscopic systems. The problem in hand may contain simultaneous ballistic and diffusive flow of phonons, which are scattered by complex defect types such as vacancies, dislocations, interfaces, or nanoinclusions. Furthermore, it may involve novel materials which have not yet been thoroughly explored.

Heat transport in bulk semiconductors is now tractable from a first-principles perspective, thanks to theoretical and computational advances in the last decade. Central to such treatments is the Boltzmann transport equation (BTE) for phonons \cite{ziman} in its space- and time-independent form. While the first practical numerical solution algorithm of the phonon BTE dates back to 1995 \cite{omini_iterative_1995,Sparavigna1,Sparavigna2}, it only achieved popularity after it was combined with density functional theory (DFT) in 2007 \cite{BroidoAPL}, giving birth to \textit{ab-initio} thermal transport calculations as they are understood today \cite{mingo_ab_2014}.

Key advantages of this kind of methods include their quantitative predictive ability, their accounting for the quantum character of phonons, and their lack of reliance on parameterized functions such as interatomic potentials or semiempirical scattering rates, in contrast with more traditional approaches like Callaway-like models \cite{callaway_model_1959} or classical molecular dynamics simulations \cite{schelling_comparison_2002}. Examples of successful applications of \textit{ab-initio} thermal conductivity calculations in recent years are too numerous to cite here, and range from technologically mainstream 3D semiconductors \cite{ma_intrinsic_2016} to graphene \cite{lindsay_flexural_2010} and other innovative 2D materials \cite{zeraati_highly_2016}. They have been used to provide physical insight about known behavior \cite{delaire}, to critically analyze other models \cite{ma_examining_2014} and experimental data \cite{carrete_low_2014}, and to perform high-throughput explorations across chemical space \cite{carrete_finding_2014}. Several software packages implementing these calculations are available, including \texttt{ALAMODE} \cite{tadano_anharmonic_2014}, \texttt{phono3py} \cite{phono3py}, \texttt{PhonTS} \cite{chernatynskiy_phonon_2015}, and \texttt{ShengBTE} \cite{sheng}.

In view of the success of \textit{ab-initio} thermal transport calculations to date, it is highly desirable to extend them beyond the domain of bulk systems to the multiscale situations described above. Unfortunately, when a dependence in position is included in the Boltzmann transport equation one faces the so-called ``curse of dimensionality'' \cite{peikert_wavelet_2011} because of the six spatial dimensions involved in the problem (3D real space + 3D reciprocal space). A direct approach to solving the BTE as a linear system, in the same spirit as for homogeneous bulk materials, is unfeasible because of the associated explosion in matrix sizes. Traditional Monte Carlo solvers, which have proven very useful to escape the ``curse'' in other domains of materials science, are of limited use in the context of the phonon BTE \cite{peraud_monte_2014}. The reason is that in most situations of practical interest phonons are fairly close to equilibrium, meaning that Monte Carlo methods spend most of the computational effort on characterizing the already well known equilibrium component, even though it is only the small deviation from the equilibrium distribution that matters for thermal transport. This makes deviational Monte Carlo approaches \cite{vrmc1} a much better choice. The idea behind them is treating the equilibrium component of phonon distributions in an analytical fashion, and devoting all computational resources to simulating the non-equilibrium component by splitting it into an arbitrary number of particle-like packets whose stochastic trajectories can be traced in space and time.

In this work we describe \texttt{almaBTE}, a collection of solvers of the Boltzmann transport equation for phonons in space- and time-dependent settings, supplemented by a set of auxiliary programs that make up a consistent framework for studying phonon-mediated heat transport in structured semiconducting systems starting from first principles. The 1.0 release of \texttt{almaBTE} features a new and improved implementation of all the functionality of \texttt{ShengBTE} \cite{sheng} for bulk systems, specialized models for the in-plane and cross-plane thermal conductivity of thin films \cite{apl-thinfilms}, a variance-reduced Monte Carlo solver of the BTE for the steady-state regime in one dimension, a general implementation of the superlattice models first presented in Ref. \onlinecite{chen_role_2013}, and an analytical solver for 1D models introduced in Ref. \onlinecite{prb-levy1}.

The present paper is structured as follows. In section \ref{sec:general} we describe the general features of \texttt{almaBTE} as a software package and how its components are articulated. Then, in section \ref{sec:programs} we introduce each program in the package along with a brief summary of the physical and mathematical formalism behind it. We include a set of examples of application in section \ref{sec:examples}. Finally, we summarize our main conclusions in section \ref{sec:conclusions}.

\section{General structure of \texttt{almaBTE}}\label{sec:general}
\myfig[!htb]{width=\columnwidth}{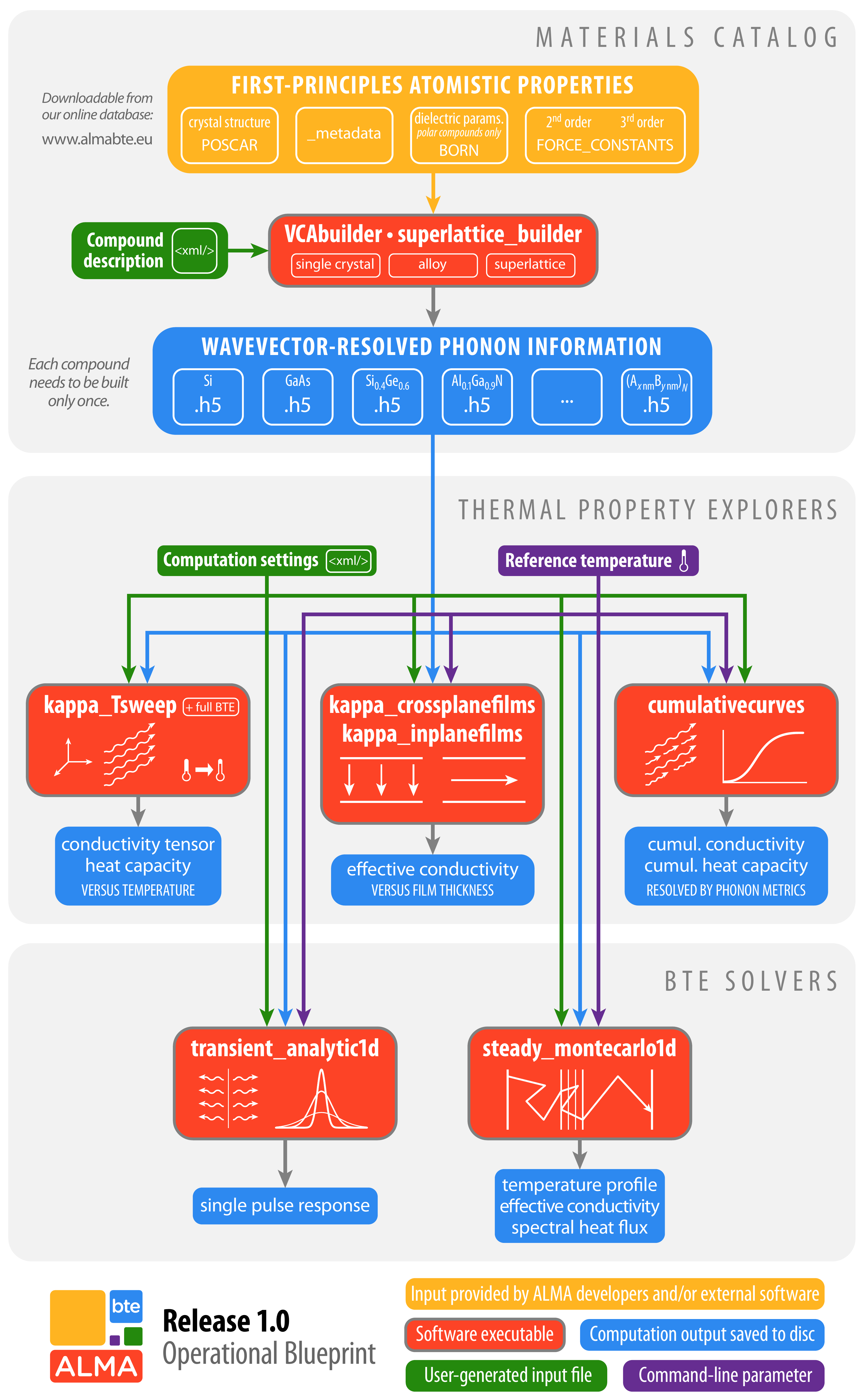}{ALMA 1.0 blueprint}
Figure \ref{blueprint} illustrates the overall fashion in which the different pieces of \texttt{almaBTE} are organized. A from-scratch calculation starts from the top of the diagram, with the creation of a catalog of ``effective crystals'' that can be used as building blocks of more complex structures. In this context, an ``effective crystal'' can be assigned an atomic structure and a phonon spectrum. 

In terms of implementation, the information about a material is contained in an HDF5 file \cite{hdf5} with a filesystem-like structure comprising several HDF5 groups (analogous to directories):

\begin{itemize}
\item The \texttt{/crystal\_structure} group contains the lattice parameters, atomic positions and chemical information of the material.
\item The \texttt{/qpoint\_grid} group stores information about the phonon frequencies, group velocities and wave functions on a regular $n_a\times n_b\times n_c$ grid in reciprocal space.
\item The \texttt{/scattering} group can either be empty or consist of an arbitrary collection of subgroups, each of them accounting for a particular kind of elastic scattering. Temperature-independent scattering rates arising from a breakdown of periodicity are stored here. As of version 1.0 of \texttt{almaBTE}, there are two datasets stored in this subgroup for each superlattice profile, as described below. Scattering by other kinds of crystallographic defects has already been implemented \cite{katre_unraveling_2016} and will be rolled out as part of future releases.
\item The \texttt{/threeph\_processes} group contains a list of allowed three-phonon absorption and emission processes along with their amplitudes. Since intrinsic phonon scattering is assumed to be dominated by these processes, this is the essential ingredient required to compute the intrinsic phonon scattering rates at any temperature.
\end{itemize}

Currently, HDF5 files can be generated for three kinds of materials: single crystals, alloys treated in the virtual crystal approximation (i.e, treated as statistical averages of several single crystals) and superlattices described as the combination of an alloy and a set of barriers (see the next section). As can be seen in the scheme in Fig.~\ref{blueprint}, creating an HDF5 file requires input from \textit{ab-initio} calculations, notably a geometrical description of each compound and the second- and third-order derivatives of its potential energy at equilibrium. The format of the input files is similar to that used in ShengBTE \cite{sheng}, with the addition of a new \texttt{\_metadata} file containing parameters such as supercell sizes. Moreover, an XML input file for the HDF5 builders needs to be written to instruct the programs as to how to combine the inputs to build a description for a single material. In the case of an alloy this entails providing a set of concentrations, while a superlattice can be described in terms of the composition of each of its layers. Examples of all required input files can be found in the \texttt{almaBTE} distribution. In order to promote collaboration and avoid duplication of work, we have started a repository of publication-quality input files from \textit{ab-initio} calculations at \url{http://www.almabte.eu}. The database currently covers about twenty compounds of scientific and technological interest. Users are encouraged to use the files in their own research and to contribute descriptions of new compounds.

The HDF5 files can be fed to the remaining programs, classified as ``thermal property explorers'' and ``BTE solvers'' in Fig.~\ref{blueprint}, and described in detail below. Each of them requires its own XML input file, providing it with a description of the operations to be carried out. The output is stored in text-based formats that are both readable and easily parseable by common data analysis software.

\texttt{almaBTE} has been designed with portability in mind. The code is written in standard C++11 making heavy use of the \texttt{Boost} libraries, and using  \texttt{Eigen} for linear algebra operations. The build system uses \texttt{CMake} to take care of dependencies automatically. API and user documentation are included with the code, along with a suite of unit tests built using the  \texttt{Google Test} library. The code is regularly built and tested with the \texttt{CLang} compiler suite on Linux and macOS, and with \texttt{g++} on Linux. Furthermore, we have built a containerized version using \texttt{Docker} that can be used on an even wider range of platform, including Microsoft Windows. It is also available for download from the web site.

The performance-critical regions of the code have been parallelized using MPI through the \texttt{Boost::MPI} wrappers. Several central parts of the algorithms are ``embarrassingly parallel'' by nature, and can be distributed over processors with minimal need for communication. Specifically, the calculations of phonon frequencies and wave functions, the search for allowed three-phonon processes and the calculation of scattering matrix elements can be efficiently parallelized over wavevectors. In non-I/O-constrained calculations, the performance of \texttt{almaBTE} has been found to scale almost linearly at least up to $512$ cores.

\section{Programs included in the package}\label{sec:programs}

\subsection{VCAbuilder}

This binary is a general-purpose builder of HDF5 files for single crystals and alloys treated in the virtual crystal approximation (VCA).

For a single crystal, \texttt{VCABuilder} reads the structural parameters (lattice vectors, elements and positions) from a POSCAR file in the format employed by the \texttt{VASP} \cite{vasp_general_1, vasp_general_2, vasp_general_3, vasp_general_4} density functional theory (DFT) package. Next, it reads the set of second derivatives of the potential energy of a supercell with this crystal structure from a FORCE\_CONSTANTS\_2ND file in the format employed by the \texttt{phonopy} \cite{phonopy} lattice dynamics software. Those elements are also known as harmonic, or second-order, interatomic force constants (IFCs). It then creates a regular, $\Gamma$-centered $n_a \times n_b \times n_c$ grid in reciprocal space and, for each point in the grid, builds the dynamical matrix based on the atomic positions and harmonic IFCs. If the compound under study contains polar bonds, this information can be complemented with values of its dielectric tensors and the Born effective charges \cite{spaldin_beginners_2012} of each atom. Those are read from a BORN file similar to the ones used by \texttt{phonopy}, and used according to the mixed-space algorithm proposed by Wang \textit{et al.}  \cite{wang_mixed-space_2010, wang_polar_2016}. By diagonalizing the corrected dynamical matrix at each wavevector, the program obtains a set of allowed vibrational frequencies and a corresponding set of phonon wave functions describing the polarizations of each atom in each mode.

\texttt{VCAbuilder} then performs a search for allowed three-phonon processes among all the vibrational modes on that regular grid. Let us use generalized indices $k, k'\ldots$ here and in the remaining of this work to label both the wavevectors and the polarizations (branches) of phonons. Two kinds of three-phonon processes are possible: absorption (+) processes where two phonons $k$ and $k'$ coalesce into a single phonon $k''$ , and emission (-) processes, where a single incident phonon $k'$ scatters into two outgoing phonons. For a three-phonon process to be allowed, it must preserve both energy and momentum. Conservation of momentum is trivially enforced on a regular grid, simply by performing the search only among phonons for which $k \pm k' = k''$, where the equality must be understood in a modular sense, i.e., taking into account the periodicity of reciprocal space. In contrast, angular frequencies $\omega$ are not uniformly distributed, meaning that in general modes satisfying the condition $\omega_k \pm \omega_{k'} = \omega_{k''}$ for the conservation of energy will not be sampled by the grid. The solution adopted in \texttt{almaBTE} is based on a linear extrapolation of angular frequencies around each grid point based on the calculated group velocities, as described in Ref. \onlinecite{PhysRevB.85.195436} and previously implemented in \texttt{ShengBTE} \cite{sheng}. In practice this translates into a parameter-free locally adaptive density estimation scheme with a Gaussian kernel. However, users have the possibility to artificially reduce the width of each Gaussian by a constant factor in order to speed up the calculation. Although this shortcut often leads to very little loss of accuracy, this should always be carefully checked.

For each allowed three-phonon process, the program then computes a scattering amplitude using the formulas in Ref.~\onlinecite{sheng}. The essential ingredient is a set of third-order derivatives of the potential energy of a supercell of the crystal, i.e., the third-order or anharmonic IFCs. Those are read from a FORCE\_CONSTANTS\_3RD in the same format as used by \texttt{ShengBTE}.

The number of allowed three-phonon processes for a typical grid is very high ($10^6-10^{12}$) which leads to high requirement of memory and CPU time. Such processes searching is optimized by using the rotations from the space group of the crystal to determine the quotient group of the wavevectors in the grid. A smaller set of wavevectors, which we term the ``irreducible wavevectors'' is then built by taking an arbitrary representative from each equivalence class in the quotient group. To do this we rely on the information about the space group of the crystal provided by \texttt{spglib}. We then restrict the search to three-phonon processes in which the first phonon ($k$) has an irreducible wavevector. Any property can be extrapolated from the irreducible set to the full wavevector grid by using the rotation operations and taking into account its tensor character (scalar, vector, etc.). In particular, relaxation times are scalar values, and the phonon populations can be parameterized in terms of a vector quantity $\boldsymbol{F}_k$ as detailed in Ref. \onlinecite{sheng}.

In order to build a description of an $n$-ary alloy, \texttt{VCABuilder} needs the same set of input files (POSCAR, \_metadata, FORCE\_CONSTANTS\_2ND, FORCE\_CONSTANTS\_3RD and possibly BORN) for each of the $n$ components. Furthermore, each of those components must contain the same number of elements, with identical stoichiometry and the same crystal structure, so that a one-to-one mapping can be established between the crystallographic sites of each structure. \texttt{VCABuilder} computes effective parameters for a virtual crystal representing the alloy, using simple arithmetic averages $\phi_{\mathrm{VC}}=\sum_i x_i \phi_i$, where $x_i$ is the mole fraction of component $i$ (normalized so that $\sum_i x_i = 1$), and $\phi_i$ can stand for a lattice vector, an atomic coordinate, a second- or third-order IFC, the dielectric tensor or a Born effective charge. This relatively crude approximation has the advantage of not requiring any further ab-initio calculations beyond those for the pure components of the alloy, no matter what the concentrations are, and has been shown to afford reasonable results \cite{li_thermal_2012}. On the other hand, its limitations must be borne in mind as it does not account for correlations, local relaxations or changes in electronic structure due to alloying.

The compositional disorder present in alloys increases the probabilities of elastic phonon scattering. This is taken into account by treating the alloy as a random mass perturbation upon the reference virtual crystal. Elastic scattering amplitudes are computed using the formula derived by Tamura \cite{tamura_isotope_1983}, in complete analogy to mass disorder scattering in a single crystal. As described below, the ingredients in this treatment are the phonon spectrum of the virtual crystal and the standard deviation of the mass at each crystallographic site.

\subsection{superlattice\_builder}

This component of \texttt{almaBTE} can create HDF5 files containing descriptions of the phonon spectrum and scattering properties of binary superlattices. An idealized superlattice consist of a repeated alternating sequence of two different compounds, $A$ and $B$, whose lattices need to be reasonably well matched to prevent dislocations. Additionally, \texttt{superlattice\_builder} requires that the structures of the two compounds satisfy the conditions of the virtual crystal approximation described above. A period of the superlattice is specified by a number of layers $N_{\mathrm{layers}}$ and a set of mole fractions $\left\lbrace x_i\right\rbrace_{i=1}^{N_{\mathrm{layers}}}$ describing the concentration of component $A$ in each layer, with component $B$ present, correspondingly, in mole fractions $\left\lbrace 1 - x_i \right\rbrace_{i=1}^{N_{\mathrm{layers}}}$. For an ideal "digital" profile, this consists of either 0 or 1 values for each layer. Segregation during growth leads to composition profiles that can differ markedly from the digital profile. A crucial difference between the digital profile and the segregated one is that the former is completely homogeneous in the directions parallel to the SL layers, whereas the latter contains local compositional disorder in all three spatial directions. This interplay between 1D and 3D compositional variation has an important effect on phonon scattering.

We model a superlattice as a periodic perturbation upon a reference virtual crystal with mole fraction $\sum\limits_i x_i / N_{\mathrm{layers}}$ of component $A$. The perturbation contains two contributions, from mass disorder and from barriers. The former accounts for the random distribution of components $A$ and $B$ within each layer of the superlattice, and is modeled using Tamura's formula with the compositions $x_i$ and $1 - x_i$ in each layer $i$. The latter depends only on the coordinate along the growth direction, and represents the effect of the average mass profile scattering due to nanostructuring. This methodology was first introduced in Ref.~\onlinecite{chen_role_2013} and validated by reproducing experimental thermal conductivity measurements on Si/Ge superlattices.

The barrier contribution to scattering comes from an intense enough perturbation as to require a fully converged treatment in the framework of perturbation theory. More specifically, the perturbed phonon wave functions are different enough from the unperturbed ones as to cause a breakdown of Fermi's golden rule (or, equivalently, the Born approximation). Since the scatterer is translationally invariant in the two directions parallel to the superlattice layers, phonons scattered by barriers preserve the parallel components of their wavevectors: an incident state can only be scattered to other states with the same parallel wavevector. Hence it is convenient to use a hybrid real/reciprocal space representation: the direction perpendicular to the superlattice layers is treated in real space, whereas the translationally invariant directions parallel to the layers are considered in reciprocal space.  The contribution to the total elastic scattering rates from barriers is obtained using the optical theorem:

\begin{equation}
  \tau^{-1}_{\mathrm{barrier}} = -\frac{1}{\omega_{b}\left(q_{\perp}, \boldsymbol{q}_{\parallel}\right)}\Im\left\lbrace\mel{q_{\perp}, \boldsymbol{q}_{\parallel}, b}{\boldsymbol{t}^{+}\left(\boldsymbol{q}_{\parallel}\right)}{q_{\perp}, \boldsymbol{q}_{\parallel},b}\right\rbrace.
  \label{eqn:optical}
\end{equation}

\noindent Here, $\boldsymbol{q}_{\parallel}$ stands for the conserved parallel component of the wavevector, $q_{\perp}$ for the component perpendicular to the layers, $b$ for the phonon branch index, and $\boldsymbol{t}^+$ for the causal t matrix, obtained from the causal Green's function $\boldsymbol{g}^+$ as:

\begin{equation}
  \boldsymbol{t}^+ \left(\boldsymbol{q}_{\parallel}\right) = \left[\mathbf{1} - \boldsymbol{V}\boldsymbol{g}^+ \left(\boldsymbol{q}_{\parallel}\right) \right]^{-1} \boldsymbol{V},
  \label{eqn:causalt}
\end{equation}

\noindent where $\boldsymbol{V}$ is the perturbation matrix describing the barrier. Note that, in the spirit of the virtual crystal approximation, \texttt{superlattice\_builder} models the perturbation as a periodic change in mass and neglects the possible change in force constants. Each component of the causal Green's function involves an integral in the non-conserved component of the wavevector:

\begin{align}
  &g^+_{\left(i, \alpha\right), \left(j, \beta\right)} \left(\boldsymbol{q}_{\parallel}, \omega^2\right) =\\\nonumber &\lim\limits_{\epsilon\rightarrow 0^+} \sum\limits_b \frac{1}{\mathrm{length}\left[L\left(q_{\perp}\right)\right]}\int\limits_{L\left(q_{\perp}\right)}
  \frac{\braket{i,\alpha}{q_{\perp}, \boldsymbol{q}_{\parallel},b}\braket{q_{\perp}, \boldsymbol{q}_{\parallel},b}{j,\beta}}{\omega^2 - \omega^2_b\left(q_{\perp}, \boldsymbol{q}_{\parallel}\right)+i\epsilon}
  d q_{\perp}.
  \label{eqn:1dgf}
\end{align}

\noindent The integration region $L\left(q_{\perp}\right)$ is the segment in reciprocal space determined by the intersection of the reciprocal-space unit cell and the line passing through $\left(q_{\perp},q_{\parallel}\right)$ in the supercell growth direction. In contrast to the 2D and 3D cases, this 1D Green's function contains drastic divergences at any point where the branches do not have a continuous derivative. This includes the physical divergences at the band edges, but also artifactual ones wherever two linear segments of a band meet with different slopes. Hence, an approach based on linear interpolation of the bands between each pair of points in a grid (analogous to the tetrahedron method \cite{lambin_computation_1984} in 3D) becomes numerically problematic as the number of grid points is increased. \texttt{superlattice\_builder} implements a new approach to this integral in the analytically correct $\epsilon\rightarrow 0^+$ limit. Since the group velocities are also available during the calculation, we build a cubic interpolation of each band, with continuous derivatives throughout the whole $L\left(q_{\perp}\right)$ segment. The contributions to the Green's function from each sub-segment in the regular grid can still be expressed analytically; the increased mathematical complexity is more than compensated by the drastically reduced number of points that need to be included in the grid since the piecewise cubic polynomial provides a much better approximation to the real bands than a set of linear segments. Detailed formulas are provided in Appendix~\ref{app:cubic}.

After computing the mass-disorder and barrier contributions to scattering, \texttt{superlattice\_builder} stores them as two separate subgroups of the \texttt{/scattering} HDF5 group. A unique suffix is generated for each superlattice, so that several different profiles can be stored in the same file.

\subsection{kappa\_Tsweep}
This executable provides efficient computations of the thermal conductivity $\kappa(T)$ versus ambient temperature. Unlike \texttt{ShengBTE}, the allowed three-phonon emission/absorption processes and associated scattering matrix elements $V_{k \, k' \, k"}^{\pm}$ do not need to be recomputed since in \texttt{almaBTE} this information is stored in an HDF5 file (see previous sections). This renders the evaluation of thermal bulk properties over a large number of ambient temperatures a fairly quick task. After reading the HDF5 file from disk, the program precomputes the total temperature-independent 2-phonon scattering rates $1/\tau_{\text{2ph}}$. All that remains is then to  compute the 3-phonon scattering rates at each of the temperature values using the formulas detailed in Ref. \onlinecite{sheng}, obtain the total scattering rates $1/\tau = 1/\tau_{\text{2ph}} + 1/\tau_{\text{3ph}}$, and evaluate the bulk thermal conductivity as described below.

One of the contributions to $1/\tau_{\text{2ph}}$ comes from mass-disorder scattering, either due to the presence of several isotopes in the crystal or to alloying \cite{tamura_isotope_1983}:

\begin{equation}
  \tau^{-1}_{k, \mathrm{m.d.}} = \frac{\pi \omega_k^2}{2}\sum\limits_{k',i} \frac{\sigma^2\left(m_i\right)}{\left\langle m_i \right\rangle^2} \left\vert\sum\limits_{\alpha}\braket{i, \alpha}{k}\braket{k'}{i, \alpha}  \right\vert^2 \delta\left(\omega_k - \omega_{k'}\right).
  \label{eqn:tamura}
\end{equation}

\noindent Here, $i$ runs over crystallographic sites in a single unit cell, and $\left\langle m_i \right\rangle$ and $\sigma^2\left(m_i\right)$ are the arithmetic mean and the variance of the distribution of masses at site $i$, respectively.

In case additional sources of elastic scattering, e.g.~dislocations, are present the corresponding datasets can be read from the input HDF5 file and added on top of mass-disorder scattering to obtain the total $1/\tau_{\text{2ph}}$. The case of superlattices is special, since mass-disorder scattering is computed in a layer-by-layer fashion. Hence, for superlattices the result of Eq. \eqref{eqn:tamura} is replaced by, and not simply added to, the elastic scattering rates read from the file.

\subsubsection{Relaxation time approximation}
Under the relaxation time approximation (RTA), the component of the thermal conductivity tensor along Cartesian axes $\alpha$ and $\beta$ is immediately obtained as

\begin{equation}
  \kappa_{\alpha,\beta} = \sum \limits_{k} C_k \, \frac{v_{\alpha,k}v_{\beta,k}}{\left\vert \boldsymbol{v}_k \right\vert} \, \Lambda_k,
\end{equation}

\noindent In this equation, $C_k$ is the mode contribution to $C(T)$, the volumetric heat capacity, $\boldsymbol{v}$ the group velocity, and $\Lambda(T) = \left\vert\boldsymbol{v}\right\vert \, \tau(T)$ the mean free path. The sum over $k$ must be interpreted as the combination of a sum over branches and an average over the Brillouin zone. The same convention applies for the remainder of the manuscript.

\subsubsection{full BTE computations}
It is also possible to solve the linearized BTE for bulk media subjected to a temperature gradient while fully accounting for all scattering terms. This is achieved through a linear system of equations (described in detail in Ref. \onlinecite{sheng}) that takes the form

\begin{equation}
  \boldsymbol{A} \boldsymbol{F} = \boldsymbol{B} \label{kappa_fullBTE_system}
\end{equation}

\noindent whose solution comprises a Cartesian vector $\boldsymbol{F}_k = \left(F_{k,x}, F_{k,y}, F_{k,z} \right)$ for each phonon mode $k$. Those vectors act as generalized mean free paths in the expression of the thermal conductivity tensor:

\begin{equation}
  \kappa_{\alpha,\beta} = \sum \limits_{k}^{} C_k \, v_{k, \alpha} \, F_{k, \beta}
  \label{eqn:beyondrta}
\end{equation}

\noindent In \texttt{ShengBTE}, the system (\ref{kappa_fullBTE_system}) contained unknowns for every point in the wavevector grid and was solved iteratively, starting from the RTA solution $\boldsymbol{F}_{k,\alpha} = \tau_k \cdot \boldsymbol{v}_{k,\alpha}$ as initial guess. Here, we exploit rotational symmetries to rewrite the system solely in terms of unknowns pertaining to irreducible grid points, and then solve this system directly with \texttt{Eigen} routines.

\subsection{kappa\_crossplanefilms, kappa\_inplanefilms}
These executables enable efficient assessment of 1D thermal transport in thin films for both in-plane ($\parallel$) and cross-plane ($\perp$) configurations. The effective RTA thermal conductivity in a film of thickness $L$ is evaluated as

\begin{equation}
  \kappa_{\text{eff}}(L) = \sum \limits_{k} S_k(L) \, C_k \, \| \boldsymbol{v}_k \| \, \Lambda_k \, \cos^2 \vartheta_k
\end{equation}

\noindent Here $S$ is a ``suppression function'' that accounts for the additional phonon scattering induced by the film boundaries and $\vartheta$ is the angle between the group velocity and transport axis. It is important to note we evaluate the wavevector-resolved $S$ on a mode-by-mode basis and thereby carefully account for any crystal anisotropies. This can be particularly important when analyzing non-cubic crystals. Most literature resolves $S$ by phonon frequency under the assumption of isotropic phonon dispersions.

\subsubsection{In-plane transport}
The in-plane geometry is described by two Cartesian vectors: the film normal $\boldsymbol{n}$, and an orthogonal vector $\boldsymbol{u}$ along which the thermal transport is to be evaluated (Fig. \ref{inplanefilm_geometry}a).
\myfig[!htb]{width=0.3\textwidth}{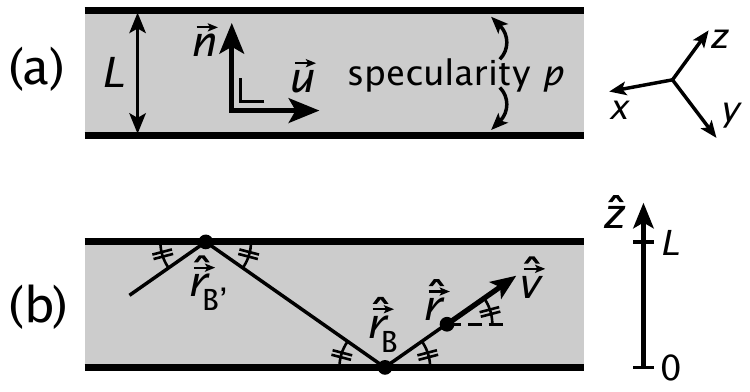}{In-plane thin film transport: (a) geometric configuration, (b) phonon trajectory involved in evaluating conductivity suppression function.}

The film boundaries are assumed to possess the same specularity $0 \leq p \leq 1$ for all phonon frequencies/wavelengths. For algebraic convenience, we carry out our computations in a transformed coordinate system in which the film normal is aligned with the new $z$ axis. This is easily achieved by performing a 3D rotation of the original coordinates with rotation matrix
\begin{equation}
  \boldsymbol{R} = \begin{bmatrix} \cos \phi \cos \theta & \sin \phi \cos \theta & -\sin \theta \\ - \sin \phi & \cos \phi & 0 \\ \cos \phi \sin \theta & \sin \phi \sin \theta & \cos \theta \end{bmatrix}
\end{equation}
in which $\phi$ and $\theta$ are respectively the azimuthal and polar angles of the film normal in original coordinates, i.e. $\boldsymbol{n}/\left\vert \boldsymbol{n} \right\vert = \left(\cos \phi \sin \theta , \sin \phi \sin \theta , \cos \theta\right)$. Quantities expressed in transformed coordinates will be marked with a $\hat{\,\,\,}$ symbol; for example, the rotated phonon group velocities read $\hat{\boldsymbol{v}} \equiv \boldsymbol{R} \boldsymbol{v}$. The suppression factor $S$ of an individual phonon mode now follows from the same reasoning that underpins the familiar Fuchs-Sondheimer formalism, albeit before any integrations over solid angle have been applied. Specifically, working backwards from Eq. (11.5.3) in Ref. \onlinecite{ziman} we have
\begin{equation}
  S_{\parallel} = \frac{1}{L} \int \limits_{0}^{L} \left[ 1 - \frac{(1-p) \, \exp \left( - \| \hat{\boldsymbol{r}} - \hat{\boldsymbol{r}}_B \| / \Lambda \right)}{1 - p \, \exp \left( - \| \hat{\boldsymbol{r}}_B - \hat{\boldsymbol{r}}_{B'} \| / \Lambda \right)} \right] \mathrm{d}\hat{z} \label{Sinplane_integral}
\end{equation}
The meaning of the various points along the phonon trajectory is illustrated in Fig. \ref{inplanefilm_geometry}b. We obtain for any $\hat{v}_{\hat{z}} \neq 0$
\begin{eqnarray}
  \frac{\| \hat{\boldsymbol{r}} - \hat{\boldsymbol{r}}_B \|}{\Lambda} & = & \frac{\mathrm{sgn}(\hat{v}_{\hat{z}}) \, \hat{z} + \frac{1}{2} [1-\mathrm{sgn}(\hat{v}_{\hat{z}})] L}{K \cdot L} \\
  \frac{\| \hat{\boldsymbol{r}}_B - \hat{\boldsymbol{r}}_{B'} \|}{\Lambda} & = & \frac{1}{K}
\end{eqnarray}
where we introduced the effective Knudsen number \cite{ziambarasa_phonon_2006}
\begin{equation}
  K = \frac{(\hat{v}_{\hat{z}} / \| \boldsymbol{v} \|) \cdot \Lambda}{L} = \frac{\hat{\Lambda}_{\hat{z}}}{L}
\end{equation}
The integration (\ref{Sinplane_integral}) yields
\begin{equation}
  S_{\parallel} = \frac{1 - p \, \exp \left(- \frac{1}{K}\right) - (1-p) \, K \, \left[1 - \exp \left(- \frac{1}{K}\right)\right]}{1 - p \, \exp \left(- \frac{1}{K}\right)} \label{Sinplane}
\end{equation}
Notice that $S_{\parallel}(K \rightarrow 0) = 1$ regardless of the specularity $p$. This confirms recovery to bulk transport in very thick films ($L \rightarrow \infty$), and correctly signals that phonons which never interact with the film boundaries ($\hat{v}_{\hat{z}} = 0$) contribute their full nominal conductivity regardless the film thickness.
\subsubsection{Cross-plane transport}
The cross-plane configuration can be fully specified by a single Cartesian vector, since the transport is evaluated along the film normal: $\boldsymbol{u} \equiv \boldsymbol{n}$. Here the film boundaries are considered to act as perfectly absorbing black bodies. The suppression function can be compactly written as
\begin{equation}
  S_{\perp} = \frac{1}{1 + 2 K} \label{Scrossplane}
\end{equation}
where the Knudsen number can be simply evaluated as $K = \Lambda \, |\cos \vartheta|/L$ without the need for coordinate transforms. Although (\ref{Scrossplane}) is not rigorously exact, we have shown in prior work \cite{apl-thinfilms} that this suppression function reproduces Monte Carlo simulations of the effective conductivity within a few percent and is closely compatible with semi-analytic solutions of the BTE in a finite domain. \texttt{almaBTE} can furthermore perform an accurate parametric fitting of the computed $\kappa_{\perp}(L)$; details on these compact models are available in Ref. \onlinecite{apl-thinfilms}.

\subsection{cumulativecurves}
This executable reveals the contributions of phonon modes to bulk heat capacity and thermal conductivity. In particular, it computes curves of the form
\begin{equation}
  C_{\Sigma}(X^{\ast}) = \sum \limits_{X_k \, \leq \, X^{\ast}}^{} C_k \;\; , \;\; \kappa_{\Sigma}(X^{\ast}) = \sum \limits_{X_k \, \leq \, X^{\ast}}^{} C_k \, \| v_k \| \, \Lambda_k \, \cos^2 \vartheta_k
\end{equation}
resolved by phonon parameter $X$ which can be mean free path $\Lambda$; relaxation time $\tau$; ``projected'' mean free path $\Lambda |\cos \theta|$, which as we just saw plays a central role in cross-plane film transport; frequency $\nu$; angular frequency $\omega = 2 \pi \nu$; or energy $E = h\nu = \hbar \omega$.

\subsection{transient\_analytic1d}
The \texttt{analytic1d} module in \texttt{almaBTE} enables exploration of 1D time-dependent phonon transport in infinite bulk media. Specifically, the module provides semi-analytic solutions of the single pulse response of the Boltzmann transport equation under the relaxation time approximation (RTA-BTE). The heat flow geometry is taken as one-dimensional, meaning that the thermal field only depends on one Cartesian space coordinate $x$.

Let us consider a planar heat source located at $x=0$ that at time $t=0$ injects a pulse of thermal energy with unit strength 1$\,$J/m$^2$ into the medium. As is customary, we assume that the input source energy gets distributed across the various phonon modes according to their contributions to the heat capacity. The RTA-BTE describing the transient evolution of the deviational volumetric thermal energy $g(x,t)$ of a phonon mode then reads
\begin{equation}
  \frac{\partial g_k}{\partial t} + v_{x,k} \, \frac{\partial g_k}{\partial x} = - \frac{g_k - C_k \Delta T}{\tau_k} + \frac{C_k}{\sum C_k} \, \delta(x) \, \delta(t)
\end{equation}
Subscripts $x$ indicate quantities measured along the thermal transport axis. The equation needs to be complemented by a closure condition expression the conservation of energy:
\begin{equation}
  \sum \limits_{k}^{} \frac{1}{\tau_k} \, (g_k - C_k \Delta T) = 0.
\end{equation}
Analytic solutions of the stated 1D problem have been previously derived in transformed domains, first for isotropic media \cite{minnich1D} and then generalized to crystals with arbitrary anisotropy by several of the \texttt{almaBTE} developers \cite{prb-levy1}. In Fourier-Laplace domain $(x,t) \leftrightarrow (\xi,s)$ the macroscopic deviational energy density $P \equiv (\sum C_k ) \times \Delta T$ is given by
\begin{eqnarray}
  P(\xi,s) & = & \frac{\sum C_k \, \Xi_k(\xi,s)}{\sum (C_k / \tau_k) \left[ 1 - \Xi_k(\xi,s) \right]} \label{BTEsol1} \\
  \text{in which} \quad \Xi_k(\xi,s) & = & \frac{1 + s \tau_k}{(1 + s \tau_k)^2 + \xi^2 \, \Lambda^2_{x,k}} \label{BTEsol2}
\end{eqnarray}
where $\Lambda_x \equiv |v_x| \, \tau$ as usual. The \texttt{analytic1d} module computes several quantities of interest by inverting the solution (\ref{BTEsol1}--\ref{BTEsol2}) semi-analytically to real space and time, as described below.
\subsubsection{Temperature profiles $\Delta T(x,t)$}
Fourier-Laplace inversion of the thermal fields can be greatly simplified if we limit ourselves to weakly quasiballistic regimes $|s| \tau \ll 1$ for which we have
\begin{eqnarray}
  P(\xi,t) & \simeq & \exp \left[ -\psi(\xi) \, t \right] \\
  \text{with} \,\, \psi(\xi) & = & \sum \frac{C_k \, \xi^2 \Lambda_{x,k}^2}{\tau_k [1 + \xi^2 \Lambda_{x,k}^2]} \,\, \biggr / \sum \frac{C_k}{1 + \xi^2 \Lambda_{x,k}^2}
\end{eqnarray}
This solution ignores purely ballistic transport effects, but offers excellent performance at temporal scales exceeding characteristic phonon relaxation times (typically 1--10$\,$ns). To perform Fourier inversion to real space
\begin{equation}
  P(x,t) = \frac{1}{\pi} \, \int \limits_{0}^{\infty} P(\xi,t)\,\cos(\xi x) \, \mathrm{d}\xi \label{invfourier}
\end{equation}
we first execute a piecewise second-order Taylor series expansion over consecutive $\xi$ intervals
\begin{equation}
  \exp[-\psi(\xi) \, t] \simeq A_{0,n}(t) + A_{1,n}(t) \, \xi + A_{2,n}(t) \, \xi^2
\end{equation}
which can then be integrated fully analytically:
\begin{multline}
  x \neq 0 : \int (A_0 + A_1 \xi + A_2 \xi^2) \, \cos(\xi x) \, \mathrm{d}\xi = - \frac{2 A_2 \sin(\xi x)}{x^3} \\
  + \frac{(A_1 + A_2 \xi) \, \cos(\xi x)}{x^2} + \frac{(A_0+ A_1 \xi + A_2 \xi^2) \sin(\xi x)}{x} \end{multline}
Downscaling this solution by the bulk heat capacity yields the desired temperature profile: $\Delta T(x,t) = P(x,t)/\sum C_k$.
\subsubsection{Source response $\Delta T(x=0,t)$}
Experiments typically have no access to the internal thermal fields but can only probe the temperature rise generated at the heat source itself. The latter can be computed by evaluating the Fourier inversion (\ref{invfourier}) at $x=0$ using linear quadrature. Limiting ourselves again to quasiballistic regimes, we find
\begin{equation}
  P(x=0,t) \simeq \sum \limits_{n}^{} \frac{\left[ \exp (- \psi_n \, t) - \exp (- \psi_{n+1} \, t) \right] \, \Delta \xi_n}{\left( \psi_{n+1} - \psi_n \right) t}
\end{equation}
in which $\psi_j \equiv \psi(\xi_j)$ and $\Delta \xi_j = \xi_{j+1} - \xi_j$.
\subsubsection{Mean square displacement $\sigma^2(t) = \int x^2 P(x,t) \mathrm{d}x$}
The evolution of the thermal MSD, defined as the variance of the deviational energy distribution, is readily obtained in Laplace domain through moment generating properties:
\begin{equation}
  \sigma^2(s) =- \frac{\partial^2 P(\xi,s)}{\partial \xi^2} \biggr|_{\xi = 0}
\end{equation}
From the general BTE solution (\ref{BTEsol1}--\ref{BTEsol2}) we find
\begin{equation}
  \sigma^2(s) = \frac{2 \sum \kappa_k / (1+ s \tau_k)^2}{s^2 \sum C_k / \left(1 + s \tau_k\right)}
  \label{sigmasquare}
\end{equation}
where $\kappa_k = C_k \, |v_{x,k}| \, \Lambda_{x,k}$ is the phonon thermal conductivity. Note that the expression (\ref{sigmasquare}) applies to all time scales; in particular, it is also valid in purely ballistic and strongly quasiballistic transport regimes. Finally, we evaluate the time domain counterpart $\sigma^2(t)$ through numerical Gaver-Stehfest Laplace inversion \cite{inverselaplace}.

\subsection{steady\_montecarlo1d}
The \texttt{steady\_montecarlo1d} module enables Monte Carlo simulations of steady-state thermal transport in 1D multilayered structures with isothermal boundary conditions.

The computational algorithm is based on the variance-reduced deviational techniques introduced by Peraud and Hadjiconstantinou \cite{vrmc1,vrmc2,peraud_monte_2014} for numerical solution of the RTA-BTE. However, we emphasize again that our computations utilize first-principles phonon properties that are wavevector-resolved. Prior publications often employed parameterized dispersions and/or scattering rates, therefore lacking predictive power for novel materials not yet experimentally characterized. Moreover, most formulations specify phonon properties as a function of frequency, which becomes highly problematic in anisotropic crystals.

The central idea of variance-reduced methods for solving the phonon BTE is to approximate $g_k$, the nonequilibrium component of the energy distribution for each phonon mode, by the sum of a predefined number of delta functions, all of them with the same energy and each with a well defined position. The stochastic trajectories of these energy packets are then tracked in order to reconstruct $g_k$. The simulation proceeds very much like a conventional particle-based Monte Carlo, leading to the label of ``deviational particles'' for these packets. However, it is important to keep in mind that deviational particles are neither phonons nor even abstract bundles of phonons, but corrections to the analytic Bose-Einstein reference distribution. As such, a particle can be either positive or negative, depending on whether it was launched from a hot or a cold reservoir respectively.

\texttt{steady\_montecarlo1d} implements a specialized version of variance-reduced Monte Carlo that allows simulating the steady-state regime directly without the need for an equilibration phase \cite{peraud_monte_2014}. This variation on the original idea exploits the time translation invariance of the ensemble of possible trajectories due to the stationary character of steady-state phonon populations. Based on this property, deviational particles are taken to represent fixed amounts of power, instead of fixed amounts of energy.

The simulations are carried out under the linearized regime. This assumes that the temperature deviations $\Delta T$ are sufficiently small such that the phonon properties of each constituting material can be treated as location independent and equal to those evaluated at the reference temperature $T_{\text{ref}} = (T_{\text{hot}} + T_{\text{cold}})/2$. Under these conditions, deviational particles act completely independently, allowing efficient and straightforward parallellization, and their trajectories $x(t)$ can be readily evaluated in a robust way without requiring any spatial or temporal discretizations.

We now take a closer look at the various events a particle may encounter during its computational lifetime (Fig. \ref{vrmc_events}).
\myfig[!htb]{width=\columnwidth}{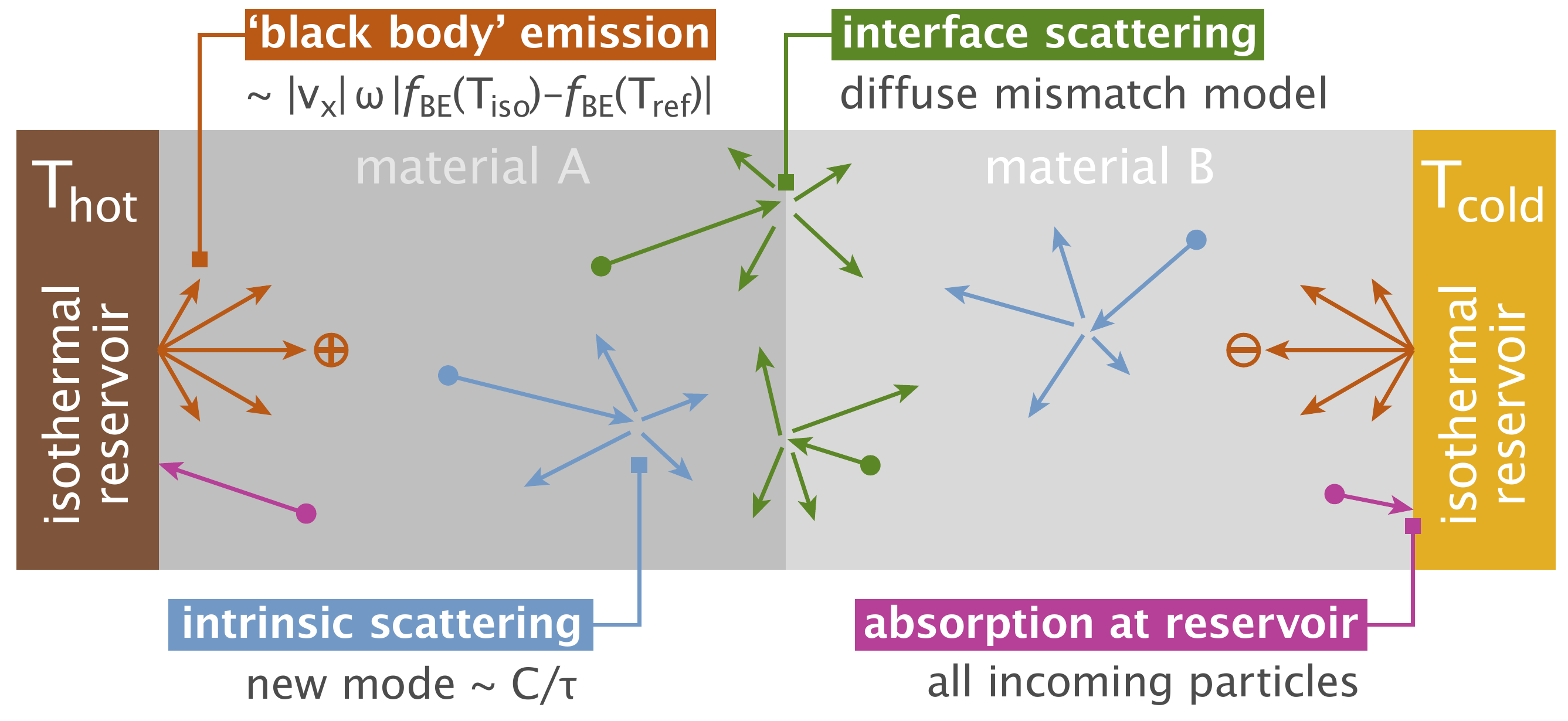}{Schematic overview of possible events encountered by deviational particles in Monte Carlo simulations.}
\subsubsection{Emission from a thermal reservoir (launch event)}
The deviational intensity [W/m$^2$] injected into the structure by an isothermal reservoir at temperature $T_{\text{iso}}$ is given by
\begin{equation}
  \mathcal{I}_{\text{iso}} = \frac{\hbar}{V} 
  \sum \limits_{k} (\boldsymbol{v}_k \cdot \boldsymbol{n})  H(\boldsymbol{v}_k \cdot \boldsymbol{n}) \omega_k \left[ f_{\text{BE}}(\omega_k,T_{\text{iso}}) - f_{\text{BE}}(\omega_k,T_{\text{ref}})\right]
\end{equation}
where $H$ is the Heaviside function and $\boldsymbol{n}$ denotes the outwardly pointing normal vector. Each particle randomly selected to be emitted from the hot or cold reservoir with respective probabilities $p_{\text{hot}} = |\mathcal{I}_{\text{hot}}|/(|\mathcal{I}_{\text{hot}}| + |\mathcal{I}_{\text{cold}}|)$ and $p_{\text{cold}} = 1-p_{\text{hot}}$, while the associated phonon mode is drawn at random with probability proportional to $(\boldsymbol{v}_k  \cdot\boldsymbol{n})  H(\boldsymbol{v}_k\cdot  \boldsymbol{n})  \omega_k  \left\vert f_{\text{BE}}(\omega_k,T_{\text{iso}}) - f_{\text{BE}}(\omega_k,T_{\text{ref}})\right\vert$. The process is illustrated by orange arrows in Fig.~\ref{vrmc_events}. The $+$ and $-$ labels on particles emitted from the hot and cold reservoirs, respectively, denote the sign of their contributions to the deviational energy.
\subsubsection{Advection steps}
A provisional travel time $\chi$ is drawn from an exponential distribution with mean $\tau_k$, and the particle provisionally moved over $\Delta x = \chi v_x$. If the advection step does not traverse any layer boundaries, the particle is moved to the new location and undergoes intrinsic scattering (blue in Fig.~\ref{vrmc_events}). Otherwise, the particle is moved to the boundary, its travel time adjusted accordingly, and undergoes either interface scattering (green) or reservoir absorption (pink). 
\subsubsection{Intrinsic scattering}
A new phonon mode is drawn with probabilities proportional to $C_k/\tau_k$, after which the simulation proceeds with a new advection step.
\subsubsection{Interface scattering}
Particles that reach the interfaces between dissimilar materials $\mathcal{M}_1$ and $\mathcal{M}_2$ are transmitted/reflected according to a diffuse mismatch model that allows for elastic mode conversions. Specifically, the particle is assigned a new phonon mode drawn from distribution $(\boldsymbol{v}^M_k \cdot \boldsymbol{n})  H(\boldsymbol{v}^M_k \cdot \boldsymbol{n})  G^M(\omega_k)/(V^M  N_{\text{tot}}^M)$ where $M \in \{ \mathcal{M}_1, \mathcal{M}_2 \}$, $\boldsymbol{n}$ are normal vectors pointing away from the interface, $G$ denotes the Gaussian regularization of the energy conservation $\delta(\omega - \omega_k)$ and $N_{\text{tot}} = N_A \times N_B \times N_C \times N_{\text{branches}}$ is the total number of available phonon modes. This is again followed by a new advection step.
\subsubsection{Absorption at a thermal reservoir (termination event).}
A particle completes its trajectory upon reaching an isothermal reservoir at either end of the structure. Its contributions to the temperature profile and heat flux are evaluated with the procedures described below, after which the simulation proceeds with the launch of the next particle.
\subsubsection{Evaluation of temperature profile}
Introducing a series of consecutive bins of width $w$ enables us to evaluate a deviational temperature profile $\Delta T(x)$ at the bin centers as follows. Each particle represents a deviational intensity
\begin{equation}
  \rho = \pm \, \frac{|\mathcal{I}_{\text{hot}}| + |\mathcal{I}_{\text{cold}}|}{N_{\text{particles}}}
\end{equation}
where the sign depends on which isothermal reservoir emitted the particle as mentioned earlier. The trajectory of each of these particles consists of one or more linear segments, each of which moved the particle inside the structure from some depth $x_i$ at some time $t_i$ to a depth $x_f$ at a time $t_f$. The bin edges further divide the segments into subsegments that span lengths $|\Delta x_n|, |\Delta x_{n+1}|, \ldots$ across bins $n, n+1, \ldots$ respectively (Figure \ref{vrmc_segment}).
\myfig[!htb]{width=0.4\textwidth}{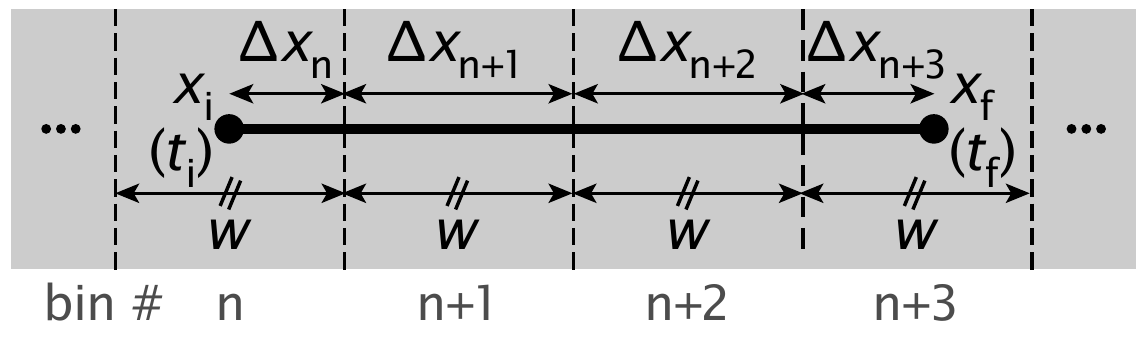}{Processing of a particle trajectory segment to evaluate its contribution to the deviational temperature profile.}
\par
The contribution to the steady-state deviational temperature in bin $l$ is now evaluated as
\begin{equation}
  \frac{1}{w  C_l}  (t_f-t_i)  \rho  \frac{| \Delta x_l |}{|x_f - x_i|}
\end{equation}
where $C_l$ is the bulk volumetric heat capacity of the material to which the $l$-th bin belongs.
\subsubsection{Evaluation of net and spectral heat flux}
Determining the net heat flux only requires observing successful particle crossings. Specifically, the heat flux in each bin is found by accumulating $\pm |\rho|/L_{\text{tot}}$ each time a trajectory segment traverses the bin center, where the $+$ sign is used for motion of positive (negative) particle moving in the hot-to-cold (cold-to-hot) direction and the $-$ sign for the opposite cases, and $L_{\text{tot}}$ is the total structure thickness. Barring stochastic fluctuations inherent to the Monte Carlo methodology, the flux profile is completely flat across the structure, due to the 1D and steady-state character of the simulation. Bin crossings can additionally be resolved in terms of the angular phonon frequency the particle was associated with at the time of the crossing, to reveal the spectral heat flux $q_{\omega}(x)$. Rather than  using simple binning with respect to $\omega$, which yields a fairly crude and noisy result, we employ locally adaptive kernel density estimation with $\Gamma$ distributions that accounts for the central frequency of the phonon mode and smoothens out the artifacts introduced by our discrete reciprocal-space grid.
\subsubsection{Evaluation of effective thermal metrics}
Once the net heat flux $q_{\text{net}}$ is known, we can easily derive the apparent thermal conductivity of the entire structure:
\begin{equation}
  \kappa_{\text{eff}} = \frac{q_{\text{net}}}{(T_{\text{hot}} - T_{\text{cold}})/L_{\text{tot}}}
\end{equation}
Alternatively, we can also express the thermal performance in terms of the effective resistivity/conductance:
\begin{equation}
  r_{\text{eff}} \equiv g_{\text{eff}}^{-1} = \frac{T_{\text{hot}} - T_{\text{cold}}}{q_{\text{net}}} = \frac{L_{\text{tot}}}{\kappa_{\text{eff}}}
\end{equation}

\section{Examples of application}\label{sec:examples}

\subsection{Thermal conductivity of bulk materials}
Figure \ref{Tsweep_results} shows computed output for diamond, Si and wurtzite GaN. Except for the former material, where proper treatment of normal scattering processes is crucial [Eq.~\eqref{eqn:beyondrta}], the RTA offers accurate conductivities within $\sim$5\% of full BTE counterparts for most common semiconductors.
\myfig[!htb]{width=\columnwidth}{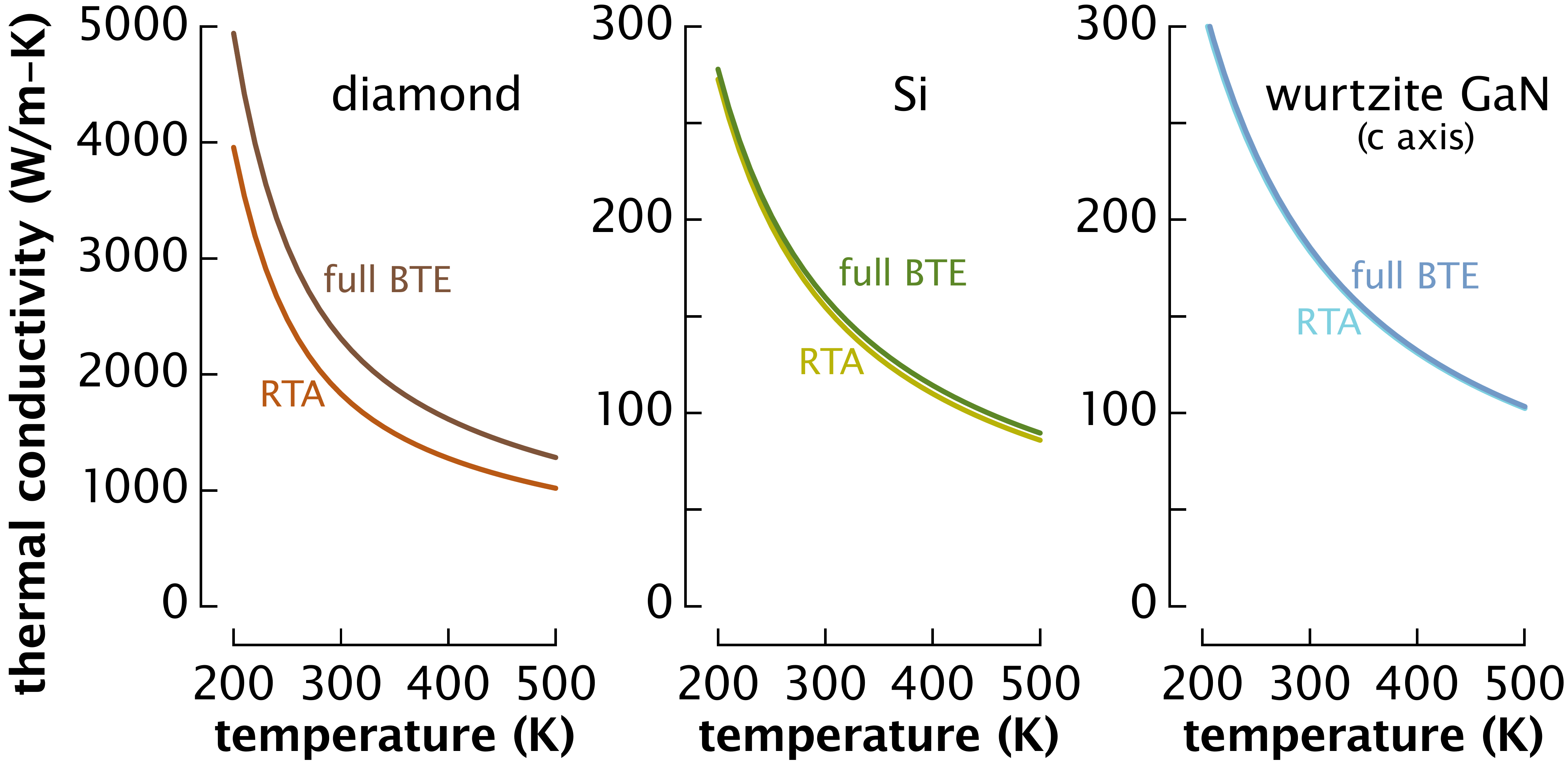}{Temperature dependence of bulk thermal conductivity.}

\subsection{Effective thermal conductivity of thin films}

The most stable phase of gallium (III) oxide (\ce{Ga2O3}) is the so-called $\beta$ phase \cite{geller_crystal_1960}, with a monoclinic structure and $10$ atoms per primitive unit cell (Fig. \ref{Ga2O3_spectrum}a shows the conventional $20$-atom cell). This relatively complex structure shows significant thermal anisotropy \cite{ga2o3} and provides a good test case for full-spectrum thin film calculations.
\myfig[!htb]{width=\columnwidth}{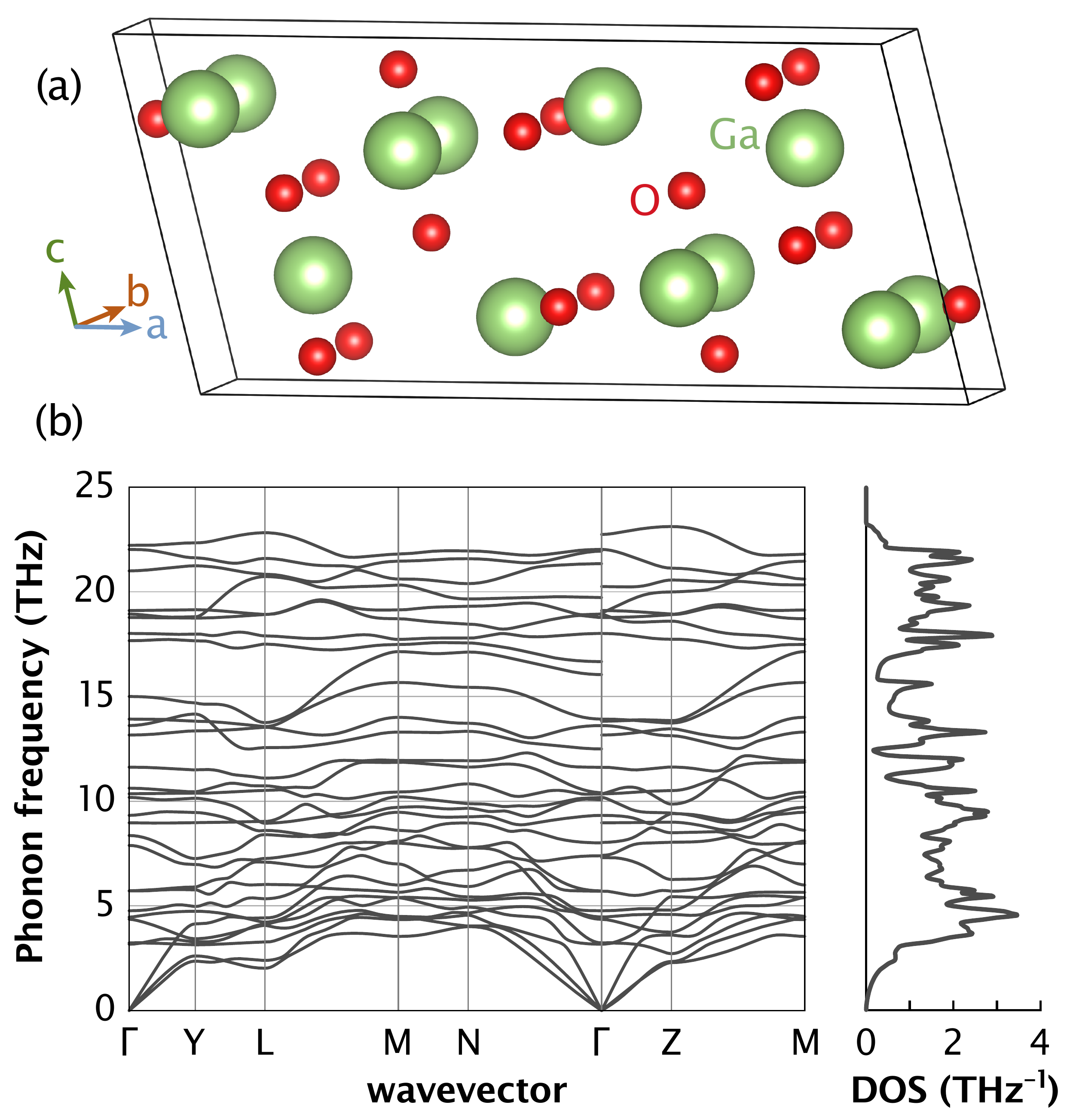}{Beta-phase gallium (III) oxide ($\beta$-Ga$_2$O$_3$) crystal. (a) unit cell, (b) computed phonon spectrum and density of states.}
\par
DFT calculations for this compound were performed with the PBEsol exchange-correlation functional as implemented in VASP. We employed the finite-temperature methodology described in Ref.~\onlinecite{ambroise_finiteT} to obtain the 2nd- and 3rd-order IFCs at 300$\,$K with a $3\times 3 \times 3$ supercell and 14th nearest neighbor cutoff, using 100 displaced configurations for each cycle, without taking into account any modification of the structure with respect to the DFT ground state. The resulting phonon spectrum and DOS is illustrated in Fig. \ref{Ga2O3_spectrum}b.
\par
Figure \ref{thinfilms_results} shows the thin film conductivities computed for a $12 \times 12 \times 12$ wavevector grid. We note that due to the low symmetry of the $\beta$-\ce{Ga2O3} crystal structure a large number of three-phonon processes must be considered, resulting in an HDF5 file more than 8 GiB in size.
\par
At room temperature, two different values of the bulk thermal conductivity along the most conductive (010) axis are now coexisting in the literature \cite{ga2o3}\footnote{We point out that the axis system in Ref. \onlinecite{ga2o3} appears to be different: the most conductive direction is designated as (110) compared to (010) in this paper and in Ref. \onlinecite{ga2o3_TDTR}.}. We find that the bulk thermal conductivity at room temperature is close to 21 W/m/K along the most conductive (010) axis. This is in agreement with the experimental measurements of Ref. \onlinecite{ga2o3_laser_flash}. In addition, we remark that the value of 27 W/m/K found in Ref. \onlinecite{ga2o3_TDTR} deviates from the general trend that can be deduced from their measurements at other temperatures. Our results are also in line with previous calculations \cite{ga2o3_ShengBTE} performed with ShengBTE \cite{sheng}, except for the (001) direction for which we find a value much closer to the experimental measurements \cite{ga2o3_TDTR} -- probably because we use a larger supercell and cutoff.
\myfig[!htb]{width=0.8\columnwidth}{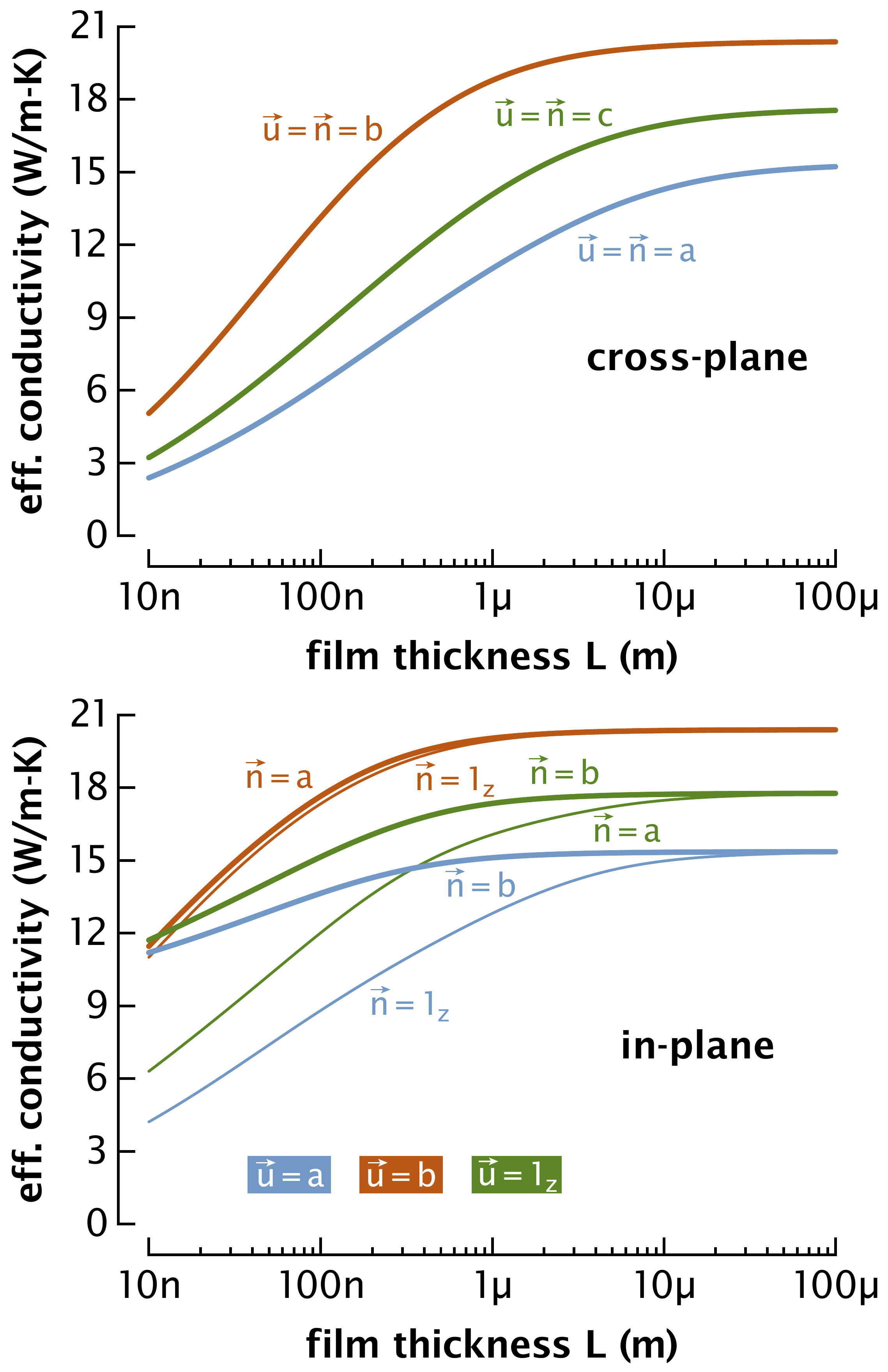}{Apparent thin film thermal conductivity in $\beta$-Ga$_2$O$_3$.}
\subsection{Cumulative thermal conductivity curves}
Figure \ref{kappacumul} shows cumulative curves resolved by mean free path and energy computed for wurtzite GaN, AlN and Al$_{0.5}$Ga$_{0.5}$N with $24 \times 24 \times 24$ wavevector grids.
\myfig[!htb]{width=0.8\columnwidth}{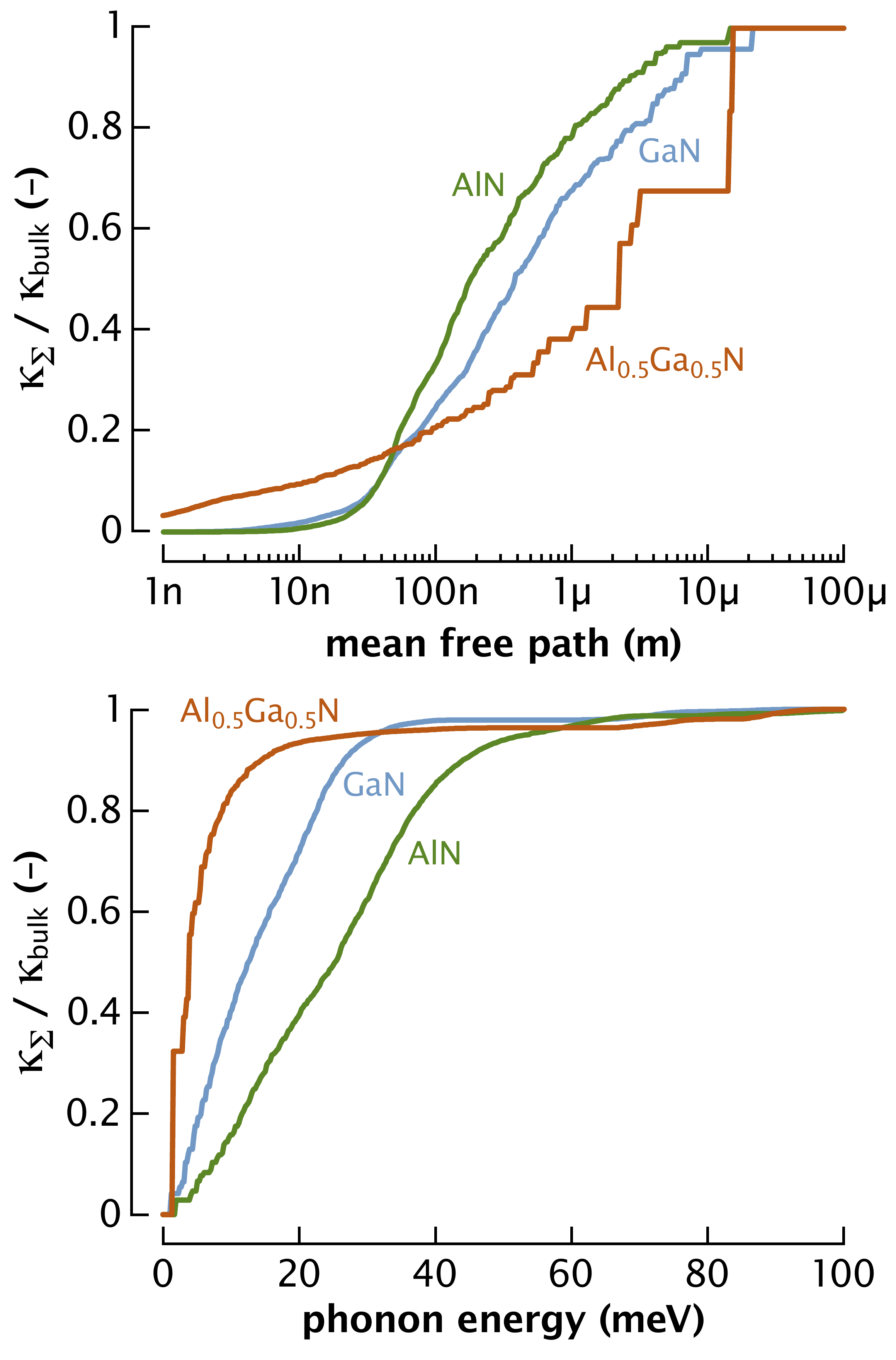}{Normalized cumulative conductivity along the c-axis for 3 wurtzite compounds.}
\subsection{Single pulse response of semi-infinite substrates}
We have computed the single pulse response over the time range 10$\,$ns--220$\,$ns in several semi-infinite substrates by doubling the \texttt{transient\_analytic1d} output results. Strictly speaking this ``method of images'' is only valid for a perfectly specular top surface. However, Monte Carlo simulations with surface specularity $p = 0.5$ show that the semi-analytic solutions are still highly adequate even for semi-diffuse boundary conditions (Fig. \ref{analytic1d_results}).
\myfigwide[!htb]{width=0.9\textwidth}{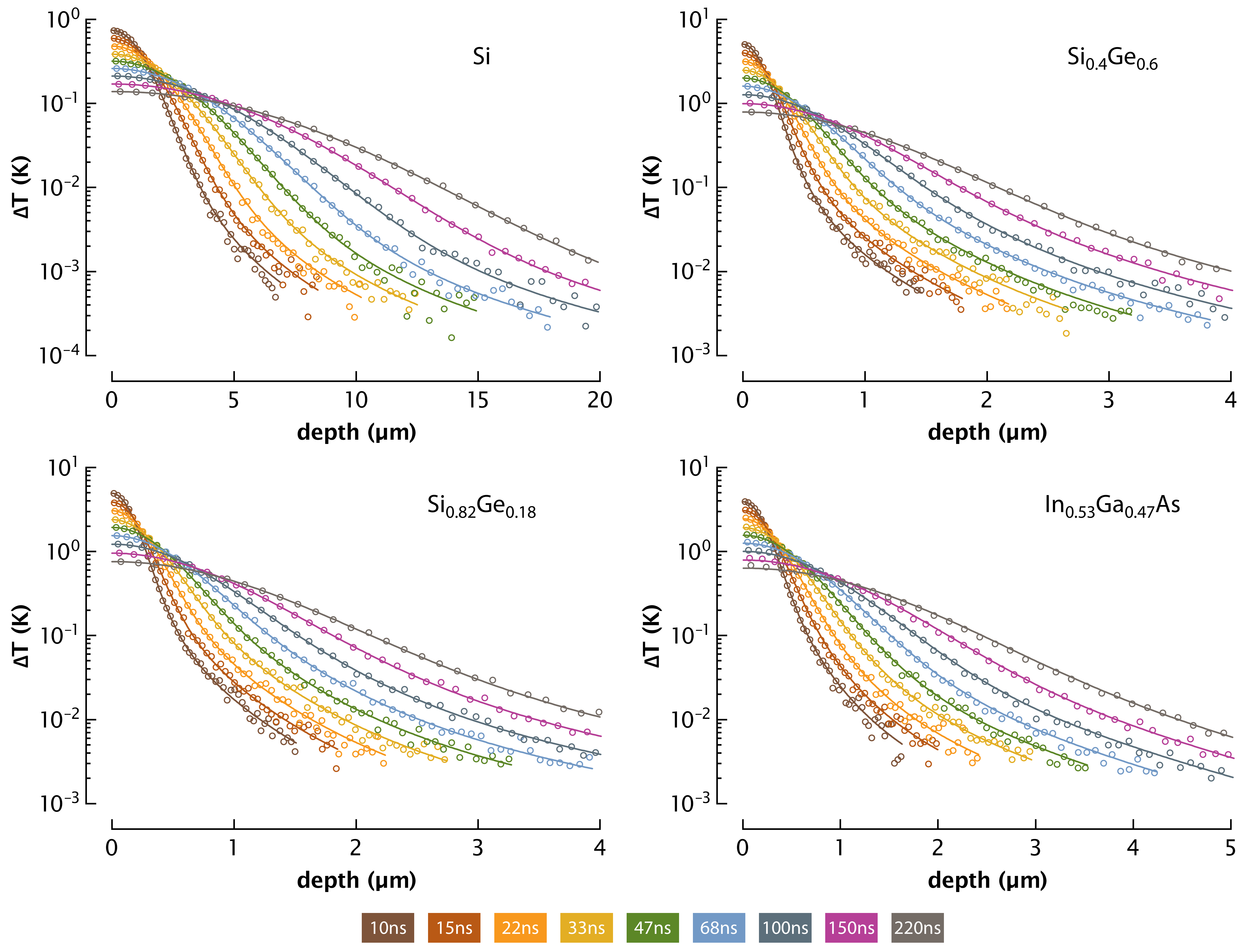}{Single pulse response in semi-infinite substrates. Semi-analytic solutions (lines) and Monte Carlo simulations (symbols) are in close agreement.}

\subsection{Steady-state Si/Ge bilayer in the steady state}
We have analyzed a Si/Ge bilayer with each layer being 200$\,$nm thick. The simulated temperature profile (Fig. \ref{vrmc_results}a) immediately reveals two phenomena that would not be encountered in a conventional Fourier diffusion treatment: the internal temperature segments are not perfectly linear, and a discontinuity appears at each of the reservoirs due to ballistic contact resistance \cite{ballisticresistance}. The substantial drop at the Si/Ge interface is also clearly visible. This interfacial thermal resistance originates in the severe phonon frequency mismatch between the two crystals, as illustrated in detail by the spectral heat flux map (Fig. \ref{vrmc_results}b). Energy conducted by optical modes in Si is reflected by the interface and must scatter to lower frequencies before it can traverse into Ge. Once inside the Ge, further scattering is visible towards the dominant acoustic modes.
\myfig[!htb]{width=0.8\columnwidth}{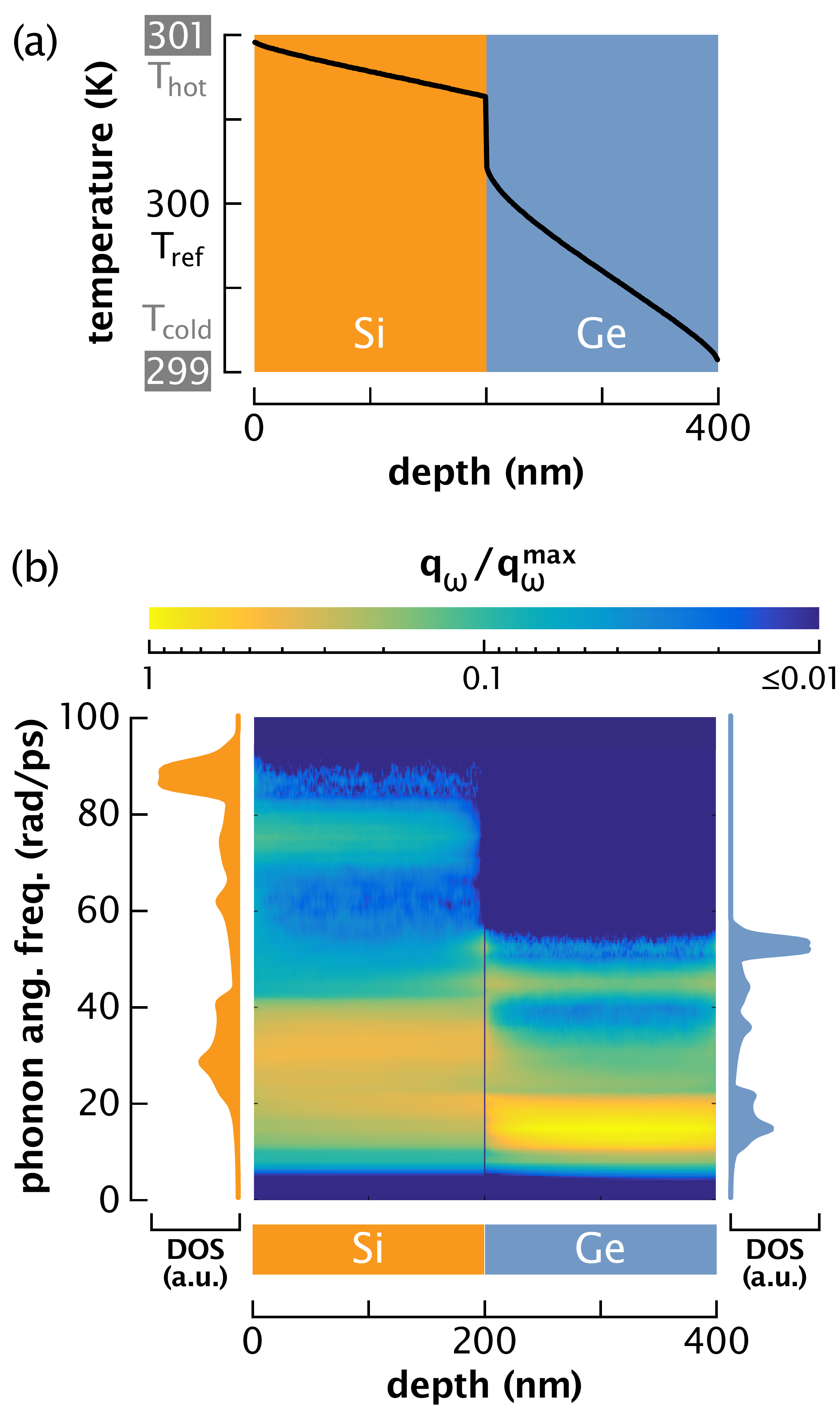}{Steady-state Monte Carlo simulation of Si/Ge bilayer: (a) temperature profile, (b) spectral heat flux.}

\section{Summary and conclusions}\label{sec:conclusions}

First-principles approaches to thermal transport have recently emerged as powerful tools to obtain predictive estimates of the thermal conductivity of crystalline semiconductors. User-friendly solvers and interfaces to first-principles codes have been developed and released to the public, helping popularize those techniques. However, this is only the first step on the way to modeling thermal transport in real semiconductors of technological interest. The conductivity of the single crystal is an upper bound to the value found in actual samples, where crystallographic defects and interfaces affect phonon scattering.

Our \texttt{almaBTE} software package provides the components needed to achieve similar predictive power across a much wider range of systems. With a modular architecture and implementing several state-of-the-art analytical and Monte Carlo approaches to the solution of the Boltzmann transport equation for phonons, \texttt{almaBTE} can provide a much richer and accurate picture of thermal transport from the nano- to the microscale than achievable with classical equations and continuum models. The code is open source and available from \url{http://www.almabte.eu} along with a database of publication-quality input files. We expect that the selected examples presented here showcase its flexibility and contribute to grow a user community around \texttt{almaBTE}.

\section*{Acknowledgments}
We thank A. Nejim, A. Bonanni, A. Rastelli, C. Giesen, C. Miccoli and N. A. Katcho for helpful discussions. This work has been supported by the European Union's Horizon 2020 Research and Innovation Programme [grant no. 645776 (ALMA)].

\begin{appendices}
 \section{Cubic interpolation approach to obtaining the 1D Green's function}\label{app:cubic}
 
 The contribution from each phonon branch to an element the 1D Green's function, evaluated at angular frequency $\Omega$, is expressed by an integral

\begin{equation}
  \mathcal{G}\left(\Omega\right) = \lim\limits_{\epsilon\rightarrow 0^+} \int_0^{2\pi} \frac{m\left(\varphi\right)}{\Omega^2 - E\left(\varphi\right) + i\epsilon} \dif\varphi.
  \label{eqn:app:gf1}
\end{equation}

\noindent Here, $\varphi\in \left[0, 2\pi\right)$ is a dimensionless parameter spanning the 1D Brillouin zone, $E\left(\varphi\right)\coloneqq \omega^2\left(\varphi\right)$ is the dispersion relation of the branch under study, and $m\left(\varphi\right)$ is a smooth function of $\varphi$. We introduce a regular partition $\varphi_n = 2\pi\frac{n}{N}$, with $n=0, 1\ldots N-1$, and at each point of the grid we sample $m_n\coloneqq m\left(\varphi_n\right)$, $E_n\coloneqq E\left(\varphi_n\right)$ and $E'_n\coloneqq\frac{dE}{d\varphi}\left(\varphi_n\right)$. In each of the subintervals we define the variable $x\coloneqq \frac{N}{2\pi}\left(\varphi - \varphi_n\right)$ and approximate the numerator and denominator of the integral by the interpolants:

\begin{subequations}\label{grp:app:interpolants}

  \begin{align}
    m\left(\varphi\right) &\simeq m_n \left(1 - x\right) + m_{n+1} x\label{eqn:app:numerator}\\
    \Omega^2 - E\left(\varphi\right) & \simeq
                                       \begin{aligned}[t]
                                         & \left(\Omega^2 - E_n\right) - E'_n x +\\
                                         + &\left(3 E_n + 2 E'_n - 3 E_{n+1} + E'_{n+1}\right)x^2 +\\
                                         + &\left(-2E_n - E'_n + 2 E_{n+1} - E'_{n+1}\right)x^3.\\
                                       \end{aligned}\label{eqn:app:denominator}
\end{align}
\end{subequations}

These approximations allows us to recast Eq. \eqref{eqn:app:gf1} as a weighted sum of the sampled values of $m\left(\varphi\right)$, after some straightforward arithmetic manipulations:

\begin{equation}
  G\left(\Omega\right) = \sum \limits_{n=0}^{N-1} \left[\left(\varpi_n^{\left(1\right)}-\varpi_n^{\left(x\right)}\right) + \varpi_{n-1}^{\left(x\right)}\right] m_n,
  \label{eqn:app:weighted}
\end{equation}

\noindent where indices must be interpreted cyclically, \textit{i.e.}, $\varphi_{-1} \equiv \varphi_{N-1}$. The partial weights $\varpi_n^{\left(1\right)}$ and  $\varpi_n^{\left(x\right)}$ are computed as rational integrals:

\begin{subequations}\label{grp:app:weights}
  \begin{align}
    \varpi_n^{\left(1\right)}& \coloneqq \frac{2\pi}{N}\lim\limits_{\epsilon\rightarrow 0^+}\int\limits_0^1 \frac{1}{p_3 + p_2 x + p_1 x^2 + p_0 x^3 + i\epsilon} \dif x\label{eqn:app:wl}\\
    \varpi_n^{\left(x\right)}& \coloneqq \frac{2\pi}{N}\lim\limits_{\epsilon\rightarrow 0^+}\int\limits_0^1 \frac{x}{p_3 + p_2 x + p_1 x^2 + p_0 x^3 + i\epsilon} \dif x,\label{eqn:app:wr}
  \end{align}
\end{subequations}

\noindent where the coefficients $\left\lbrace p_0, p_1, p_2, p_3\right\rbrace$ in the denominator are those in Eq. \eqref{eqn:app:denominator}. It is convenient to solve these integrals separately for the case when the denominator has three real roots and for the case when it has a single one. To that end, we define the reduced coefficients $a\coloneqq p_1 / p_0$, $b \coloneqq p_2 / p_0$ and $c\coloneqq p_3 / p_0$, and the discriminants \cite{numerical_recipes}:

\begin{subequations}\label{grp:app:discriminants}
  \begin{align}
    q\coloneqq & \frac{a^2 - 3b}{9}\label{eqn:app:q}\\
    r\coloneqq & \frac{a\left(2 a^2 - 9 b\right) + 27c}{54}.\label{eqn:app:r}
  \end{align}
\end{subequations}

\subsection{$r^2 < q^3\Rightarrow$ the denominator has three real roots}

The three roots are \cite{numerical_recipes}:

\begin{subequations}\label{grp:app:threeroots}
  \begin{align}
    x_0 &= -2 \sqrt{q} \cos\left(\frac{\theta}{3}\right) - \frac{a}{3}\label{eqn:app:threeroots:1}\\
    x_1 &= -2 \sqrt{q} \cos\left(\frac{\theta + 2\pi}{3}\right) - \frac{a}{3}\label{eqn:app:threeroots:2}\\
    x_2 &= -2 \sqrt{q} \cos\left(\frac{\theta - 2\pi}{3}\right) - \frac{a}{3},\label{eqn:app:threeroots:3}
  \end{align}
\end{subequations}

\noindent in terms of $\theta\coloneqq \arccos\left(r / q^{\frac{3}{2}}\right)$. From these we build the intermediate quantities

\begin{subequations}\label{grp:app:threevars}
  \begin{align}
    \alpha &\coloneqq -\frac{x_0 x_1 x_2}{p_3\left(x_0 - x_1\right)\left(x_0 - x_2\right)}\\
    \beta &\coloneqq -\frac{x_0 x_1 x_2}{p_3\left(x_1 - x_0\right)\left(x_1 - x_2\right)}\\
    \gamma &\coloneqq -\frac{x_0 x_1 x_2}{p_3\left(x_2 - x_0\right)\left(x_2 - x_1\right)},
  \end{align}
\end{subequations}

\noindent and obtain the real and imaginary parts of the weights as:

\begin{subequations}\label{grp:app:threeweights}
  \begin{align}
    \Re\left\lbrace \varpi_n^{\left(1\right)}\right\rbrace =& \frac{2\pi}{N} \Big[\alpha \mathcal{F}_1\left(x_0\right) + \beta \mathcal{F}_1\left(x_1\right) + \gamma \mathcal{F}_1\left(x_2\right) \Big]\\
    \Im\left\lbrace \varpi_n^{\left(1\right)}\right\rbrace =& -\frac{2\pi^2}{N} \begin{aligned}[t]\Big[&\left\vert\alpha\right\vert H_{\left(0,1\right)}\left(x_0\right) + \left\vert\beta\right\vert H_{\left(0,1\right)}\left(x_1\right) +\\& \left\vert\gamma\right\vert H_{\left(0,1\right)}\left(x_2\right)\Big] \end{aligned}\\
    \Re\left\lbrace \varpi_n^{\left(x\right)}\right\rbrace =& \frac{2\pi}{N} \Big[\alpha x_0 \mathcal{F}_1\left(x_0\right) + \beta x_1 \mathcal{F}_1\left(x_1\right) + \gamma x_2 \mathcal{F}_1\left(x_2\right) \Big]\\
    \Im\left\lbrace \varpi_n^{\left(x\right)}\right\rbrace =& -\frac{2\pi^2}{N} \begin{aligned}[t]\Big[&\left\vert\alpha x_0\right\vert H_{\left(0,1\right)}\left(x_0\right) + \left\vert\beta x_1\right\vert H_{\left(0,1\right)}\left(x_1\right) +\\& \left\vert\gamma x_2\right\vert H_{\left(0,1\right)}\left(x_2\right) \Big],\end{aligned}
  \end{align}
\end{subequations}

\noindent using the indicator function $H_{\left(0,1\right)}\left(x\right)$ (equal to $1$ if $x\in\left(0, 1\right)$, and $0$ otherwise) and the auxiliary function $\mathcal{F}_1\left(x\right)\coloneqq \log\left\vert \frac{1 - x}{x} \right\vert$.

\subsection{$r^2 \ge q^3\Rightarrow$ the denominator has a single real root}

The value of the root can be computed as:

\begin{equation}
  x_0 = A + B - \frac{a}{3}\text{, where}\label{grp:app:oneroot}
\end{equation}

\begin{subequations}\label{grp:app:onefirst}
  \begin{align}
    A &\coloneqq -\sgn\left(r\right) \sqrt[3]{\left\vert r\right\vert + \sqrt{r^2 - q^3}}\\
    B &\coloneqq \begin{cases}
      0 & \text{if} A=0\\
      \frac{q}{A} & \text{otherwise}.
    \end{cases}
  \end{align}
\end{subequations}

\noindent Again we introduce some intermediate quantities:

\begin{subequations}\label{grp:app:onesecond}
  \begin{align}
    \zeta &\coloneqq  A + B + \frac{2}{3}a\\
    \eta &\coloneqq \left(\frac{\zeta}{2}\right)^2 + \frac{3}{4}\left(A - B\right)^2\\
    \alpha &\coloneqq -\frac{x_0\eta}{p_3\left[x_0\left(x_0 + \zeta\right) +  \eta\right]}\\
    \beta &\coloneqq -\left(x_0 + \zeta\right)\alpha,
  \end{align}
\end{subequations}

\noindent based on which the weights adopt the following expressions:

\begin{subequations}\label{grp:app:oneweights}
  \begin{align}
\Re\left\lbrace \varpi_n^{\left(1\right)}\right\rbrace =& \frac{2\pi}{N} \left[\alpha \mathcal{F}_1\left(x_0\right) + \frac{\beta}{\eta}\mathcal{F}_2\left(\frac{\zeta}{\eta}, \frac{1}{\eta}\right)-\frac{\alpha}{\eta}\mathcal{F}_3\left(\frac{\zeta}{\eta}, \frac{1}{\eta}\right) \right]\\
    \Im\left\lbrace \varpi_n^{\left(1\right)}\right\rbrace =& -\frac{2\pi^2}{N}\left\vert\alpha\right\vert H_{\left(0,1\right)}\left(x_0\right)\\
    \Re\left\lbrace \varpi_n^{\left(x\right)}\right\rbrace =& \frac{2\pi}{N} \begin{aligned}[t]\bigg[&\alpha x_0 \mathcal{F}_1\left(x_0\right) + \alpha \mathcal{F}_2\left(\frac{\zeta}{\eta}, \frac{1}{\eta}\right)-\\&\frac{\alpha x_0}{\eta}\mathcal{F}_3\left(\frac{\zeta}{\eta}, \frac{1}{\eta}\right) \bigg]\end{aligned}\\
    \Im\left\lbrace \varpi_n^{\left(x\right)}\right\rbrace =& -\frac{2\pi^2}{N} \left\vert\alpha x_0\right\vert H_{\left(0,1\right)}\left(x_0\right),
  \end{align}
\end{subequations}

where we have defined the functions

\begin{subequations}\label{grp:app:onefuncs}
  \begin{align}
    \mathcal{F}_2\left(x, y\right)&\coloneqq -\frac{2}{\Delta}\left[\arctan\left(\frac{x}{\Delta}\right)-\arctan\left(\frac{x + 2y}{\Delta}\right)\right]\\
    \mathcal{F}_3\left(x, y\right)&\coloneqq \frac{2 x\mathcal{F}_2\left(x, y\right) + \log \left(1 + x + y\right)}{2y}\\
    \Delta\left(x,y\right) &\coloneqq \sqrt{4 y -x^2}.
  \end{align}
\end{subequations}

\end{appendices}
  
\bibliography{bibliography}

\end{document}